\documentstyle[12pt]{article}

\newenvironment{figcap}{
     \newpage\pagestyle{empty}\thispagestyle{empty}
     \section{\ }\medskip
    \begin{list}{{\large \bf Fig.~\arabic{myeqsno}}}{
      \usecounter{myeqsno}\setlength{\rightmargin}{\leftmargin}}}{
    \end{list}}
\newcounter{myeqsno}

\begin{document}
\begin{titlepage}
\title{Reentrant Behavior in the Domany-Kinzel Cellular Automaton}
\author{H.\ Rieger$^1$, A.\ Schadschneider$^{1,2}$ und 
M.\ Schreckenberg$^1$\\
\mbox{$\quad$}\\
$^1$ Institut f\"ur Theoretische Physik\\
Universit\"at zu K\"oln\\
50937 K\"oln, Germany\\
\mbox{$\quad$}\\
$^2$ Institute for Theoretical Physics\\
SUNY at Stony Brook\\
Stony Brook, NY 11794-3840, USA}

\date{}
\maketitle
\thispagestyle{empty}
\begin{abstract}
\normalsize
\noindent
We present numerical and analytical results for a special
kind of one-dimensional probabilistic cellular automaton, the so 
called Domany-Kinzel automaton. It is shown that the phase boundary
separating the active and the recently found chaotic phase exhibits
reentrant behavior. Furthermore exact results for the $p_2$=0-line
are discussed.
\end{abstract}
\vspace*{1cm}
\end{titlepage}

Cellular automata have been an intensive research field in recent years
\cite{all,wolfram} due to their computational simplicity and the wide
range of applications in various areas. Even in one dimension a
particular probabilistic variant (Domany-Kinzel automaton) of the
originally deterministic cellular automata shows a rich phase diagram
including directed percolation and other critical phenomena
\cite{domany,kinzel}.  Only recently a new phase in this model has
been explored numerically exhibiting chaotic behavior
\cite{tsallis,tsallis2,penna}. This region of the diagram, up to a
deterministic corner-point, is not accessible to exact treatments up
to now.

Nevertheless sophisticated approximation-methods, which systematically
go beyond mean-field theory, have been applied successfully
\cite{schreck}.  In the so called tree-aproximation
\cite{derrida} one finds reentrant behavior in two directions, which
is not fully understood yet.  This phenomenon has never been observed
in numerical simulations up to now \cite{tsallis,penna}.  Therefore
one might ask, whether this reentrant behavior is a real feature of
the model or just an artifact of the tree-approximation.  This issue is the
main topic of the present paper, where we try to clearify this point
with an alternative approximation method (the cluster-approximation)
as well as with large scale Monte-Carlo simulations (up to $3\times
10^6$ sites). To state the final results already at this place: The
cluster-approximation again yields reentrant behavior in two
directions and the simulations show clear evidence for reentrance near
the tricritical point.

The model we consider is defined as follows: The Domany-Kinzel PCA
consists of a one-dimensional chain of $N$ binary variables, 
$(n_1,\ldots,n_N)$, $n_i$ taking on the values $\{0,1\}$ (empty,
occupied). All sites are updated simultaneously (i.e.\
parallel) at discrete time steps and the state of each site at time
$t+1$ depends only upon the state of the two nearest neighbors at
time $t$ according to the following rule:
\begin{eqnarray}
\lefteqn{ W(n_i\mid n_{i+1},\,n_{i-1})=} \nonumber
\\ & & \frac{1}{2}\left\{
1-(2n_i-1)\left[1-2p_1(n_{i+1}+ n_{i-1})+2(2p_1-p_2) n_{i+1}\,n_{i-1}\right]
\right\} 
\end{eqnarray}
where $W(n_i\mid n_{i+1},\,n_{i-1})$ is the (time-independent)
conditional probability that site
$i$ takes on the value $n_i$ given that its neighbors have the 
values $n_{i+1}$ and $n_{i-1}$ at the previous time step. 
$p_1$ ($p_2$) is the probability that site $i$ is occupied 
if exactly one (both) of its neighbors is (are) occupied. 
If neither neighbor is occupied, the site $i$ will also become 
empty, therefore the state with all sites empty is the absorbing 
state of the PCA.

The $(p_1,p_2)$-phase diagram, as it is known up to now, consists of three
different phases. Most of it (small enough $p_1$) is dominated by the 
{\it frozen} phase, where all initial conditions eventually lead 
into the absorbing state. With other words, the activity
\begin{equation}
a(t)=\frac{1}{N}\sum_{i=1}^N n_i(t)
\label{distance}
\end{equation}
tends to zero for $t\rightarrow\infty$ within the frozen phase.
For large enough $p_1$ one enters the $\it active$ 
phase, where, starting from a random initial condition, the system
ends up in a state with a finite density of active sites. Within this active
phase one can distinguish between a chaotic and a non-chaotic part.
This difference can be seen by starting with two slighly different (random) 
initial conditions ${\bf n}(0)$ and ${\bf n'}(0)$
subjected to the same external noise (local updating
rules). Calculating the normalized distance $d(t)$ of these two systems 
\begin{equation}
d(t)=\frac{1}{N}\sum_{i=1}^N (n_i-n_i')^2
\label{distance}
\end{equation}
during the update of the replicated systems according to the rule
displayed in equation (2) of reference \cite{schreck} one observes 
a sharp transition from the chaotic phase, characterized by
$\lim_{t\rightarrow\infty}d(t)=d_\infty>0$, to the active phase
with $d_\infty=0$ (in the following we call the active/non-chaotic
phase simply the active phase). The underlying picture is that in the
latter case the system is characterized by only one attractor, which
nevertheless depends strongly on the external noise. With other words, 
in this phase the noise (and not the initial condition) dominates the 
dynamics completely. This is not true for the chaotic phase, where
the system memorizes the initial state even after infinite time.

First we present analytical results obtained by the application of the
so-called cluster-approximation already known in different contexts
\cite{ben,schreck2} as probability path method \cite{kik} or local
structure theory \cite{guto}.  In this way we check earlier results
\cite{schreck} derived with a different approximation scheme (the
tree-approximation, see \cite{derrida}).  The problem with the
dynamical rules defined above is that one cannot write down
the probability distribution of the stationary state since
no simple detailed balance condition can be derived. Therefore,
in principle, it is necessary to solve the dynamics completely
in order to obtain the equilibrium properties. This is not
possible in general.

One way out of this dilemma is to take into account systematically all
possible correlations between $m$ neighboring sites ($m$-cluster
approximation) and to treat interactions over longer distances by
conditional probabilities. More formally, given the probability
$P(n_1,\ldots,n_m)$ for the configuration $(n_1,\ldots,n_m)$
in an $m$-cluster-approximation
the probabilitiy for configuration  $(n_1,\ldots,n_l)$ with $l>m$
is approximated to be:
\begin{equation}
P(n_1,\ldots,n_l)=P(n_1,\ldots,n_m)\prod_{i=1}^{l-m}
\tilde{P}(n_{i+1},\ldots,n_{i+m-1}\mid n_{i+m})\, .
\end{equation}
Here $\tilde{P}(n_{i+1},\ldots,n_{i+m-1}\mid n_{i+m})$ denotes the
conditional probability to find site $m+i$ in state $n_{i+m}$ given
that the $m-1$ sites to the left are in the state $(n_{i+1},\ldots,n_{i+m-1})$.
A factorisation of this kind can describe the stationary state exactly only
if the interactions extend over not more than $m$ sites. A natural choice
for the conditional probaility $\tilde{P}$ is
\begin{equation}
\tilde{P}(n_{i+1},\ldots,n_{i+m-1}\mid n_{i+m})=
\frac{P(n_{i+1},\ldots,n_{i+m})}{P(n_{i+1},\ldots,n_{i+m-1})}
\end{equation}
with
\begin{equation}
P(n_{i+1},\ldots,n_{i+m-1})=\sum_{n_{i+m}=0,1}P(n_{i+1},\ldots,n_{i+m})\, .
\end{equation}
Simple examples of one-dimensional systems which can be described
exactly by a finite value of $m$ are the $p$-spin-Ising-model where
one needs $m=p$ for the exact equilibrium distribution ($m=p=2$ being
the standard one-dimensional Ising model with next-neighbour
interactions only) \cite{cris}. Another example is the the parallel
asymmetric exclusion process where again $m=2$ leads to the exact
result for the stationary state \cite{schreck2}

The phase diagram resulting from a calculation based on the cluster
approximation with $m=2$ is shown in figure 1.  Since during one
update step according to the rules equation 1 the even (odd) sites only
depend on the odd (even) sites at the timestep before we performed two
timesteps at once to deal with sites of only one fixed parity. One
firstly observes that $m=2$ is still far from the exact solution for
the stationary state.  Unfortunately higher approximations are very
hard to obtain due to the exponentially growing number of equations to
be analysed simultaneously (especially for the distance $d(t)$ with two
replicated systems).  Furthermore even for $m=2$ the resulting
equations cannot be solved analytically with final closed expressions
but have to be iterated until one finds a fixed point of the system of
equations. In order to obtain a better localisation of the phase
boundaries we applied the same method described below to analyse the
numerical data from the Monte-Carlo-simulations.

As can be seen from the figure we find reentrant behaviour 
both in $p_1$- and $p_2$-direction comparable to the result from the tree
approximation \cite{schreck}.  It seems that the tricritical point has
moved upwards, but a detailed analysis of the results suggests that it
remains on the $p_2=0$-line. For the frozen/active-phase boundary
one can go to larger clusters with higher values of $m$.
In tabular 1 the critical values $p_1^c$ ($p_2=0$) for 
of $m\le 5$ are given:

\begin{equation}
\begin{tabular}{|l|lcl|}
\hline
1/m & &  $p_1^c$ &\\
\hline
1      &  0.5    & $\pm$ & 0.0\\
0.5    &  0.6666 & $\pm$ & 0.001\\
0.3333 &  0.723  & $\pm$ & 0.001\\
0.25   &  0.745  & $\pm$ & 0.001\\
0.2    &  0.760  & $\pm$ & 0.001\\
\hline
\end{tabular}
\end{equation}

A simple least square fit leads
to  a limiting value for $p_1^c$ of about $0.810$ which is significantly
larger than the known values from the simulations \cite{penna}.

In order to test the predictions of both approximation schemes
mentioned above we performed large-scale Monte-Carlo simulations
of the Domany-Kinzel cellular automaton with probabilities
$p_1$ and $p_2$ in the vicinity of the two end-points of the
phase boundary of the chaotic phase (i.e.: $(p_1^c,0)$ and
$(1,p_2^c)$), where reentrance could occur according to the above
calculations.

The system-sizes were up to $N=3\times 10^6$ sites with periodic
boundary conditions, and the number of iterations $t_{\rm max}$ were
maximally $10^5$.  In this way one avoids self-correlations (finite
size effects), since after $t$ updates those sites separated by a distance
smaller than $t$ are correlated. Therefore $t_{\rm max}$ has to be
smaller than $N$. By choosing $N$ much larger than $t_{\rm max}$ one
improves the statistics significantly (for obvious reasons, since one
can devide the system into many statistically independent subsystems).
Therefore no finite-size effects are present in our data (which was 
checked by comparing results for different system sizes)
and we need not to perform a (non-trivial) extrapolation the infinite
system $N\to\infty$. Furthermore the probability that the system gets
trapped by the absorbing state ($n_i=0$) after time $t$ increases
with decreasing system size. This renders the simultaneous limit
$N\to\infty$ and $t\to\infty$ to a delicate point, which we also avoid by our 
approach.

Looking at the data obtained from the simulations it turned out to be
rather unreliable to try to discriminate between two phases by looking
at the long-time limit of the order parameter (activity $a(t)$ or
distance $d(t)$). Apart from the two phase-boundaries we expect
exponential decay of $a(t)$ and $d(t)$ to their asymptotic values.
Exactly on the phase-boundary we expect the spectrum of relaxation
times to extend to infinity and thus the decay to become algebraic.
This behavior is illustrated in figure 2: the activity as a function
of time is depicted in a log-log plot for increasing values of $p_1$
($p_2=0$).  We see that below a certain value the curves are bended
downwards, whereas above this value the curves are bended upwards
reflecting exactly the behaviour explained above.  The curve just in
the middle corresponding to $p_1=0.810$ is closest (as defined
quantitatively by a least square fit) to a straight line.  To
determine $p_1^c$ more accurately we performed longer runs with larger
system sizes and depict the result in figure 3. The middle curve,
corresponding to $p_1=0.8095$, is nicely approximated by an algebraic
decay with an exponent $-0.155$. This exponent
agrees well with the universal order parameter exponent 
$\beta/\nu_{\|}=-0.157\pm0.002$ determined in reference \cite{kinzel}.

From figure 3 we determined
$p_1^c$ to be $0.8095\pm0.0005$. This is the most accurate estimate of
$p_1^c$ so far. Surprinsingly it is significantly larger than the
value $0.799\pm 0.002$ obtained with a different method but with
system sizes of around $N=640$ \cite{penna}. It seems that in the
latter reference the long transient times ($\sim 10000$) together with
the small system sizes lower the critical value due to larger
correlations in the system as known from similar systems \cite{schreck2}.
As we have mentioned above the finite size scaling analysis of small
systems is by no means straightforward and cannot be done without
further ad hoc assumptions, from which our method is free. Hence, from
our point of view, the results we quote seem to be more reliable.
Note that there is no overlap even of the error bars of the two
critical values.

In figure 4 we show the same scenario for $p_2=0.04$. By the same
arguments as above we now locate the critical value of $p_1$ (i.e.\
the value at which the transition from vanishing to finite activity
takes place) to be $0.805\pm0.002$, which is significantly lower than
$p_1^c$. For larger increasing values of $p_2$ the phase boundary
between frozen and active phase bends down monotonically to smaller
values of $p_1$ terminating at the point ($p_1=0.5$, $p_2=1$) which is
exactly known since the whole $p_2=1$-line is exactly solvable.

In figure 5 a comparison of the two curves for $d(t)$ at $p_1=0.8090$,
which is below $p_1^c$, for $p_2=0$ and $p_2=0.03$ is shown. Note that 
(0.8090,0) lies within the frozen phase. The upper curve bends upwards,
which means that (0.8090,0.03) lies within the chaotic phase. This is
indicated by the schematic phase diagram depicted in the insert of
figure 5 and which has been supported by simulations of various 
parameters $(p_1,p_2)$ in this region. The two black dots represent
the two curves shown and along the arrow connecting them one finds
clear evidence for reentrant behavior. The phase boundary of the 
chaotic phase therefore bends to the left up to values around $p_2=0.03$ 
and for larger values of $p_2$ it
bends monotonically to the right until it terminates at the point
$(1,p_2^c)$. One reason for the fact that
this phenomenon was not seen in earlier simulations is that
it is in fact a marginal effect observable only in high-precision
simulation-data. 

We also performed large scale simulations around the other endpoint of the
chaotic/active phase boundary. Here it is quite evident that the
reentrant behaviour parallel to the $p_1$-axis at $(1,p_2^c)$ is in
fact an artifact of the approximation schemes and not existent in the
actual system.

Concerning the question of the conjugate field for the order-parameter
of the chaotic phase posed in reference \cite{tsallis3} one can make on the
$p_2$=0-line exact statements. Since both the activity and the chaos
order parameter obey exactly the same evolution equations
\cite{schreck} it is easy to conclude that the conjugated fields also
should be equivalent. For the activity one chooses independent random
numbers at each site and each timestep (on the $p_2$=0-line this is
just the role of $p_1$). Accordingly one chooses for the chaotic order
parameter independant random numbers at each site and each timestep
(rule $h_1$ in ref.\cite{tsallis3}). The absorbing state now
corresponds to identical variables states in the two replicas yielding the same
update since the same noise has to be applied for equal
configurations in the two systems. This picture remains valid for
$p_2>0$ although the evolution equations are no longer identical, but the
absorbing state has the same properties. Note that on the line $p_2=0$ the 
critical exponents of the order-paramater are also the same.
If universality holds
away from this line this statement should be true also for $p_2>0$ (for
$p_2=1$ it is known, that the critical exponents are different \cite{kinzel}).

In summary we have shown in this letter that, in contradiction to
previous findings, the chaotic phase in fact shows reentrant
behaviour in the vicinity of the tricritical point as predicted by
approximative analytical methods. The effect was not seen before
since it is relatively small and large scale simulations have to be made
to detect it. On the other hand in the near of the $p_1$=1-line the
predicted reentrant behaviour is absent.

Furthermore one can see that simulations of small systems with long
transient times can lead to erroneous conclusions about the locations
of the critical point as well as the shape of the phase boundary since
it is difficult to estimate the error due to self correlations.
Therefore the error-bars in reference \cite{penna} seem to neglect
these systematic errors and should be larger (which could lead to an
agreement with our results). Finally we have seen that the conjugate
field to the chaotic order parameter can directly be identified from
the equivalence to the activity order parameter on the $p_2$=0-line.

\section[*]{Acknowledgement}

We thank G.\ Kohring, D.\ Stauffer and C.\ Tsallis for interesting
discussions. This work was performed within the SFB 341
K\"oln--Aachen--J\"ulich supported by the DFG.

\begin{figcap}
\item \label{fig1}
The phase diagram calculated via the 2-cluster-approximation.
For details see text.

\item \label{fig2}
Time-dependence of the activity for $p_2=0$ and various values 
of $p_1$. The system size is $N=10^6$. Concerning the statistical
error we observe that all runs using different random numbers
yield curves that are indistinguishable on this scale.

\item \label{fig3}
Determination of $p_1^c$: The system size is $3\cdot10^6$, $p_2=0$
and $p_1=0.8090$ (lower curve), $p_1=0.8095$ (middle curve) and $0.8100$ 
(upper curve). The straight line in the middle is the function $0.49\cdot 
t^{-0.155}$. We conclude that $p_1^c=0.8095\pm 0.0005$.

\item \label{fig4}
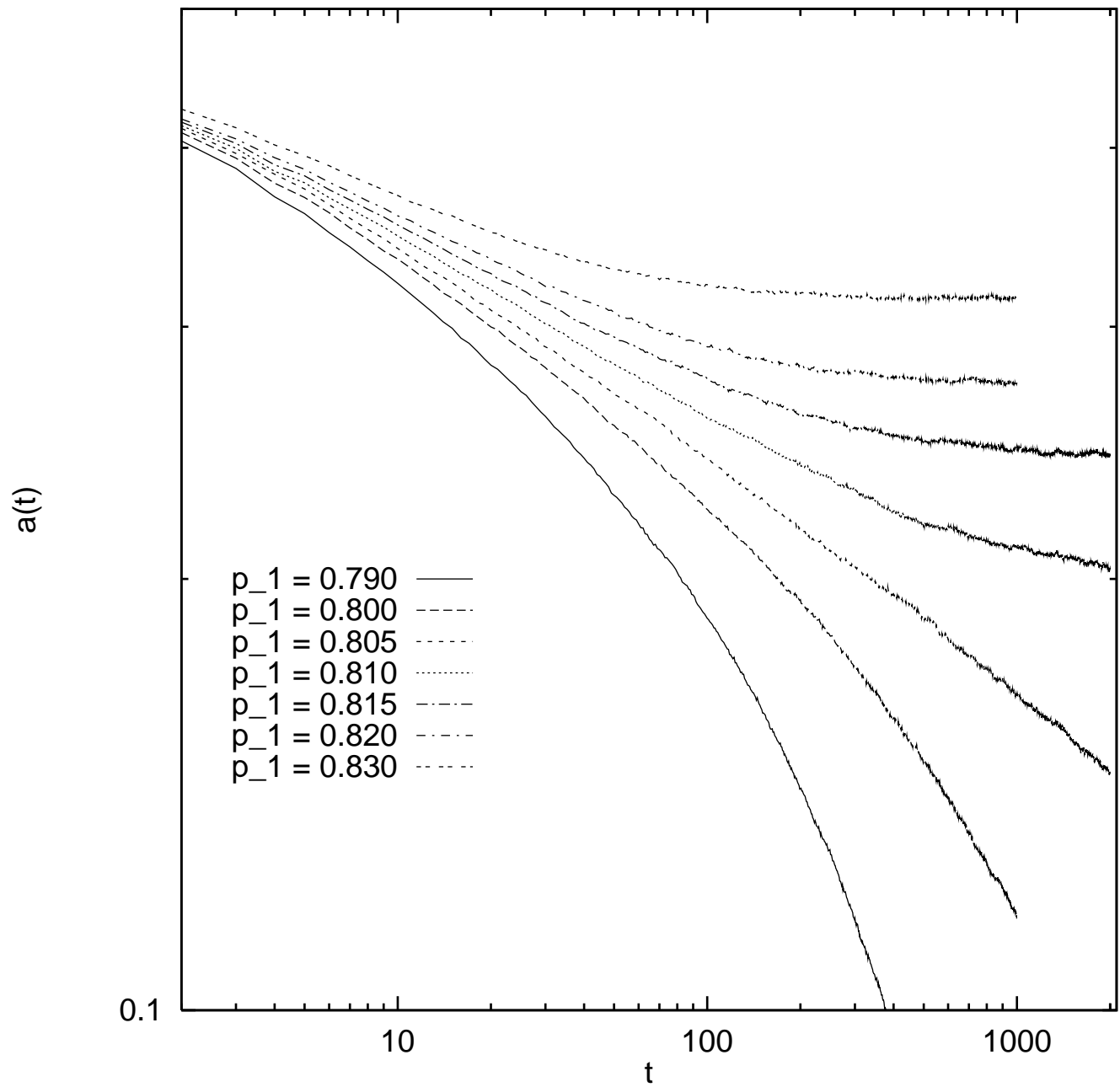
Time-dependence of the activity for $p_2=0.04$ and various values 
of $p_1$. The system size is $N=10^6$.

\item \label{fig5}
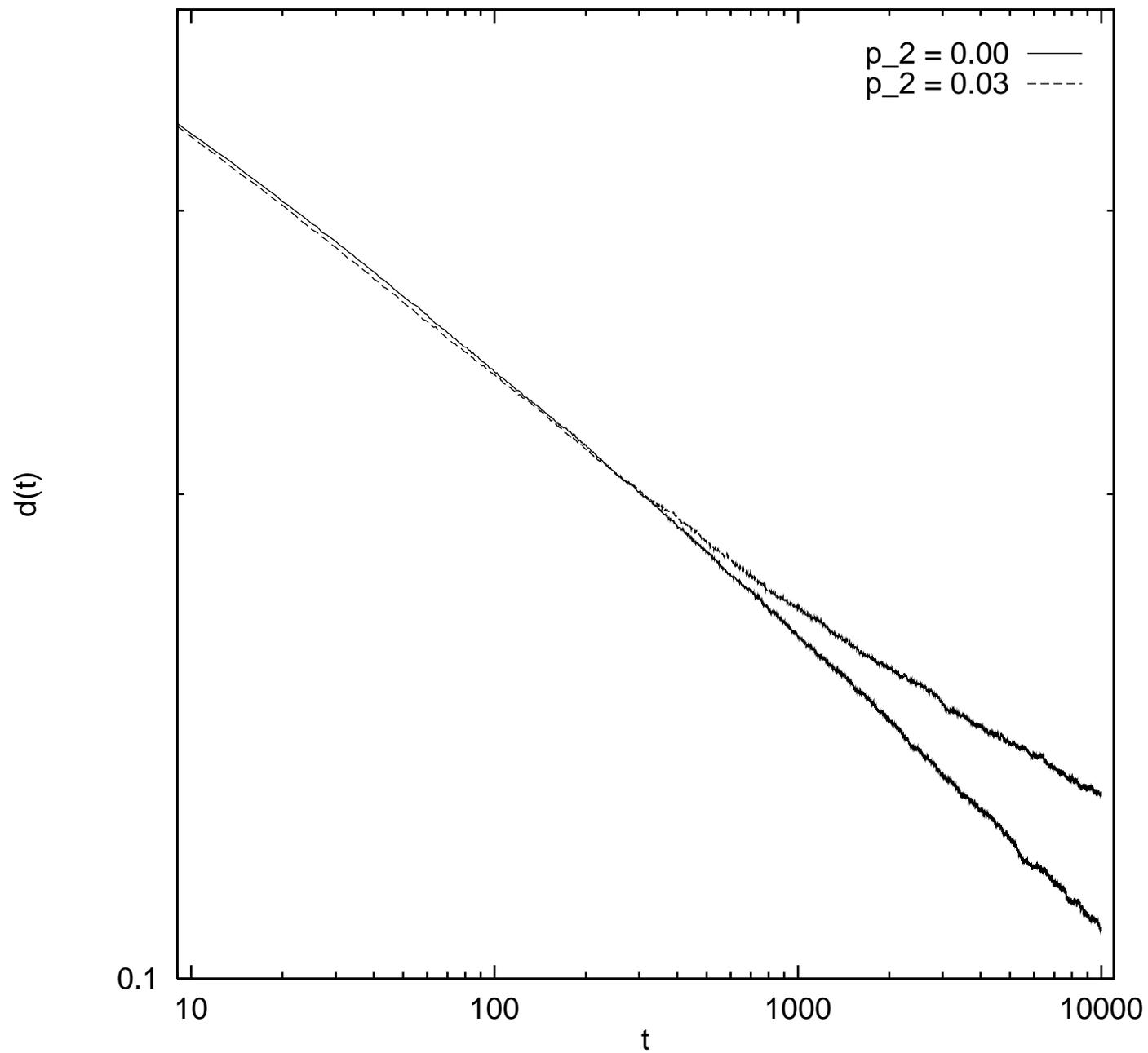
Time-dependence of the distance for $p_1=0.809<p_1^c$ and
$p_2=0.00$ (lower curve), $p_2=0.03$ (upper curve). The system size 
is $N=3\cdot10^6$. The curve for $p_2=0.03$ shows that $(0.809,0.03)$
is well inside the chaotic region, whereas $(0.809,0.00)$ lies within the 
frozen phase. This fact stronly supports reentrance.
The emerging phase diagram in the vicinity of the tricritical point
is depticted in the insert --- the two black dots represent the 
parameter values for the two curves shown.

\end{figcap}

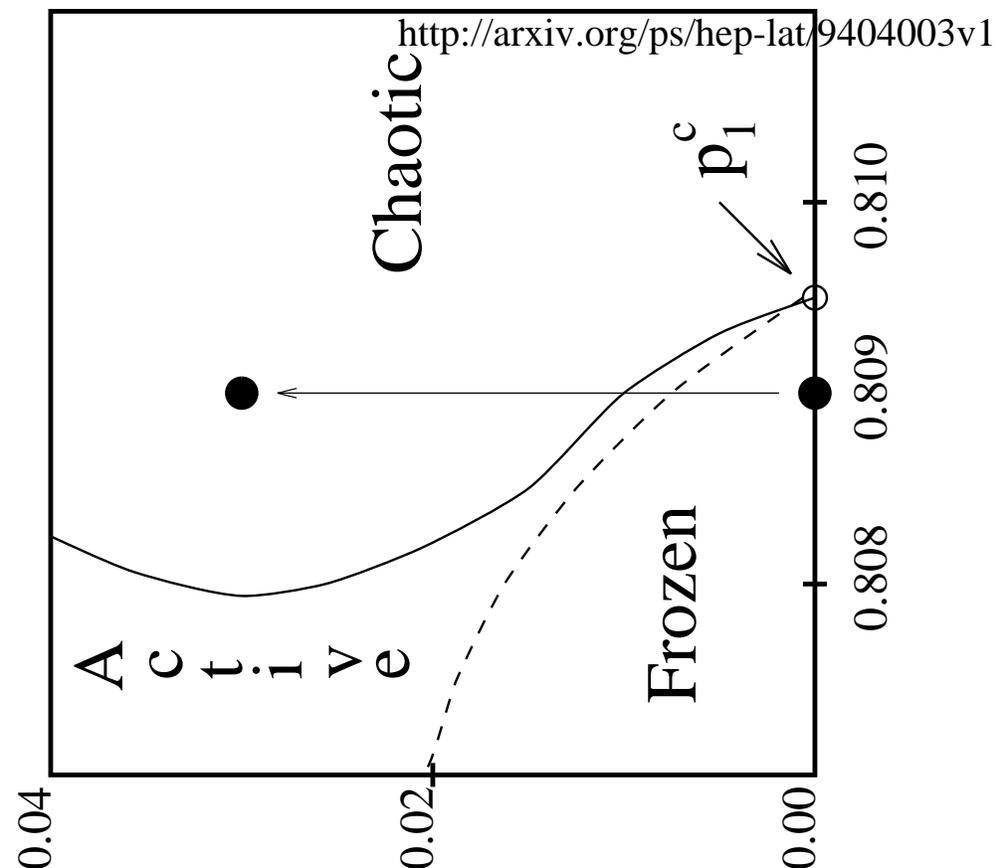
\end{document}